# Fragments, combustion and earthquakes

Oscar Sotolongo Costa[1] and Antonio Posadas Chinchilla[1,2]
1. - Cátedra de sistemas complejos "Henri Poincaré", Universidad de La Habana, Cuba
2. - Departamento de Física Aplicada, Universidad de Almería, España

Abstract
This paper is devoted to show the advantages of introducing a geometric viewpoint and a non extensive formulation in the description of apparently unrelated phenomena: combustion and earthquakes. Here, it is shown how the introduction of a fragmentation analysis based on that formulation leads to find a common point for description of these phenomena

## *I. - Introduction.*

Fragmentation is present in a plethora of natural phenomena, generally accompanying industrial, geological or other processes.
Fragmentation is responsible for the distribution of fragment sizes that appear on rocks, soils, asteroids and many other objects. Besides, its importance in biological and technological processes in evident. This is the reason for the interest of many researchers to find fragment size distribution functions (fsdf) able to fit with the distributions observed in nature.
Due to the enormous difficulties involved in the task of modeling fragmentation and fracture with the laws of mechanics, the search for a formulation in this way has been an insurmountable task.
Nevertheless, some theoretical and experimental researches have been carried out by alternative ways. Two of them can be named as "fractal formulations" (ff) [1] and "maximum entropy methods" (MEM) [2], demonstrating with their results that the methods of fractal geometry and of probability theory, together with the introduction of a new way of looking to this phenomenon, give very good and promising results.
Some statistical distributions have been proposed to fit the fsdf in a wide variety of fragmentation phenomena. Among them, it is worth to mention:

a) Weibull distribution. The cumulative number of fragments of characteristic size smaller than r can be fitted with:

$$N(r) \propto \left[1 - \exp\left(-\frac{r}{\sigma}\right)^{\alpha}\right] \qquad (1)$$

where $\alpha$ and $\sigma$ are constants.

b) Lognormal distribution. The cumulative number of fragments with size larger than r can, in other circumstances, be described as:

$$N(r) \propto \int_{r}^{\infty} \frac{\exp\left[\frac{1}{2\sigma^2}\left(\ln\left\{\frac{x-r_0}{b}\right\}\right)^a\right]}{\sigma(x-r_0)} dx \qquad (2)$$

Here, as usual, $\sigma$, $r_0$ and $b$ are constants. This equation is useful in mining processes.

c) Rossin- Ramler, very useful in problems of fuel atomization. The density of drops with radius r in a spray is usually fitted with:

$$n(r)dr \propto r^\alpha \exp\left[-\left(\frac{r}{r_0}\right)^\beta\right] dr \qquad (3)$$

d) Power law (Pareto) distribution:

$$N(r) \propto r^{-x} \qquad (4)$$

.         This distribution is of particular importance since it expresses the lack of any characteristic scale in the distribution, i.e., the smaller the fragments, the more abundant are they. The distribution has no local maximum or mean value. This has been reported in the distribution of moon craters, ice fields, islands (Korčak´s law) and many other. This is precisely the distribution that has been obtained with fractal models (See [1]). Besides. Pareto is a common behavior of a whole family of distribution functions known as "Levy distributions" whose role in physics is becoming fundamental and
 merit by itself a separate paper, since this family accomplishes the property that the behavior of a part ( In distribution, of course) is similar to that of the whole. This is a fractal property! Enough to say that many of the distribution functions used to describe fragments have Pareto as asymptotic limit.
In this article we present the theoretical and experimental results of our group in the study of fsdf in a variety of situations. Section I is devoted to the field of combustion and section II presents an attempt to apply this study in the field of earthquakes.

## *II. Fragments and combustion*

Though combustion engineers have been addict to Rossin Ramler distributions for the description of spray drops, experiments performed by our group demonstrated that the fsdf in the case of drop breaking Follows a power law. In figure 1 the experimental setup is shown

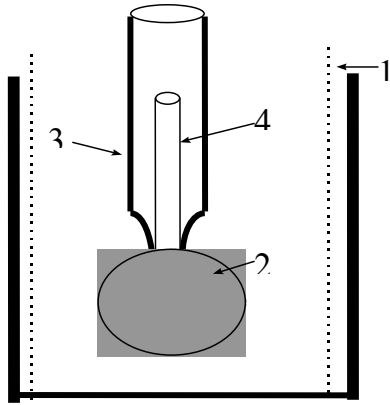

Fig. 1. Experimental setup for the fragmentation of drops:
1-transparent sheet to collect fragments, 2- oil drop. 3,4 capillaries.

The goal of the experiment was to measure fsdf of drops suffering of microexplosions, i.e., an explosive process produced in the combustion chamber when the fuel drops are emulsioned with water. Water-fuel emulsions have been used to improve the efficiency of combustion since microexplosions lead to a secondary atomization and to a larger area of interaction between fuel and oxygen during combustion. As we said, the fsdf was tested to be power law, especially when the applied pressure was high. But for lower pressures, the distribution resulted to be well fitted with a lognormal distribution. In this case a Rossin-Ramler distribution can also be used. The transition from lognormal to Pareto (For details, see [3,4]) coincides with the results reported in [5,6].

Indeed, in [5] an experiment is reported where long thin glass rods are dropped from different heights. The resulting fragments were collected ands, as the falling height increased, the fsdf transitioned from a lognormal type distribution to a Pareto. Some kind of "phase transition" was observed. The same situation was presented in [6] where the same experiment was performed but with falling mercury drops. There, a representation of fragmentation in the form of a Bethe lattice percolation process was proposed which reveals a transition from a lognormal type to a Pareto distribution in the fsdf, but the model is out of the scope of this paper. We refer the reader to [6].

It is worth to say that in the case of Pareto distribution of fuel drops, the results of the combustion are strongly dependent of the critical exponent in Eq. 4. In [3] it was shown that when $x \to 3$ in Eq. 4, the quantity of unburned matter decreases rapidly, to behave as a "perfect" combustion, with zero waste of fuel. As it was shown in [1], this corresponds to a completely self-similar distribution of drops, i.e., complete self-similarity irrespective of the scale changes made to observe the drops.

Other groups (See [2,7]) have attempted to derive the fsdf from the maximum entropy principle subject to some constraints which mainly came from physical considerations about the fragmentation phenomena> The resulting fragment distribution function describes the distribution of sizes of the fragments in a regime in which scaling is not present.

As scaling invariably occurs when the energy of the fragmentation is high, this suggests that the analysis is only applicable to low energies. However, the maximum entropy principle is universal and has an almost unlimited range of application. Consequently we would expect to be able to use it to describe the transition to scaling as the energy of the fragmentation grows.

The expression for the Boltzmann-Gibbs entropy $S$ in its Shannon form is

$$S = -k \sum_{i=1}^{W} p_i \ln p_i \qquad (5)$$

where $p_i$ is the probability of finding the system in the microscopic state $i$, $k$ is Boltzmann's constant, and $W$ is the total number of microstates.

This has been shown to be restricted to the domain of validity of Boltzmann-Gibbs (BG) statistics. This statistics describes Nature when the effective microscopic interactions and the microscopic memory are short ranged [8]. The process of violent fractioning, like that of droplet microexplosions in combustion chambers, blasting and shock fragmentation with high energies, etc, leads to the existence of long-range interactions between all parts of the object being fragmented.

Fractioning is a paradigm of non extensivity, since the fractioning object can be considered as a collection of parts which, after division, have an entropy larger than that of their union, i.e., if we denote by $A_i$ the parts or fragments in which the object has been divided, its entropy $S$ obeys $S(\bigcup_i A_i) \leq \sum_i S(A_i)$, defining a "superextensivity" in this system. This suggests that it may be necessary to use non- extensive statistics, instead of the BG statistics. This kind of theory has already been proposed by Tsallis [9], who postulated a generalized form of entropy, given by

$$S_q = k \frac{1 - \int_0^\infty p^q(x)dx}{q-1} \tag{6}$$

This integral runs over all admissible values of the magnitude $x$ and $p(x)dx$ is the probability of the system being in a state between $x$ and $x+dx$. This entropy can also be expressed as

$$S_q = \int p^q(x) \ln_q p(x) dx \tag{7}$$

The generalized logarithm $\ln_q(x)$ is defined in [8] as

$$\ln_q(x) = \frac{x^{1-q} - 1}{1-q} \tag{8}$$

where $q$ is a real number. It is straightforward to see that $S_q \underset{q \to 1}{\to} S$ recovering BG statistics.

We can maximize $\frac{S_q}{k}$. If we denote the volume of a drop by $V$ and some typical volume characteristic of the distribution by $V_m$, we can define a dimensionless volume $v = \frac{V}{V_m}$. Then, the constraints to impose are

$$\int p(v) dv = 1 \tag{9}$$

i.e., the normalization condition. The other condition we impose is mass conservation. But as the system is finite, this condition will lead to a very sharp decay in the asymptotic behavior of the dsdf for large sizes of the fragments. Consequently, we will impose a more general condition, like the "q-conservation" of the mass, in the form:

$$\int v p^q(v) dv = 1 \tag{10}$$

which reduces to the "classical" mass conservation when $q = 1$.
By the method of Lagrange multipliers, the dsdf in terms of volume is, then,

$$p(v)dv = C(1 + (q-1)\alpha_2 v)^{-\frac{1}{q-1}} dv \tag{11}$$

with

$$C = \frac{q-1}{q} \alpha_1^{\frac{1}{q-1}} \tag{12}$$

and $\alpha_1$ and $\alpha_2$ are the Lagrange multipliers.

This is a dsdf expressed as a function of the volumes of the droplets. It is convenient to formulate the problem in terms of the radius $r = v^{1/3}$:

$$f_q(r) = 3Cr^2 \left[1 + (q-1)\alpha_2 r^3\right]^{-\frac{1}{q-1}} \qquad (13)$$

The range of admissible values of q is $1 < q < 2$. This range of values of q also guarantees the adequate power law behavior since the asymptotic behavior is

$$f_q(r) \square \frac{1}{r^{\alpha+1}} \qquad (14)$$

where $\alpha = 3\frac{2-q}{q-1}$. Also, if $q \to 1$, Eq. 11 leads to:

$$f(r) = 3r^2 \exp(-r^3) \qquad (15)$$

which is the well-known Rossin-Ramler distribution.

Then, with this formulation the behavior of fragments in the process of breaking is reproduced. Thus we confirm that BG statistics cannot be applied to all fragmentation regimes and Tsallis entropy can be used to describe the transition to scaling. In this respect, the parameter , which determines the "degree of non extensivity" of the statistics, ca be related to an effective temperature of breakage.

## *II. –Fragments and earthquakes*

The Gutenberg-Richter law has motivated a mass of research. This is due to the importance of the knowledge of the energy distribution of earthquakes and its physical and practical implications. Some famous models like those of Burridge and Knopoff [10] or Olami et al [11] have focused in the mechanical phenomenology of earthquakes through simples images which capture essentials of the nature and genesis of a seism: the relative displacement of tectonic plates or the relative motion of the hanging wall and footwall on a fault and also the existence of a threshold for a catastrophic release of energy in the system. The irregular geometry of the profiles of the tectonic plates and fault planes was highlighted, for example, in De Rubeis et al [12] using a geometric viewpoint to deduce the power law dependence of earthquake energy distribution with good results. The importance of a geometric viewpoint for this phenomenon has been also highlighted in Herrmann et al [13], where an idealized representation of the fragmented core of a fault (gouge) was presented. Today it is widely accepted that most earthquakes are originated by relative motion of fault planes, whereas the images to model this energy release are diverse. The standard earthquake picture usually assigns the cause of an earthquake to some kind of rupture or some stick-slip mechanism in which the friction properties of the fault play the determinant role. A review of these viewpoints and some generated paradoxes can be found in [14]. Herrmann et al (1990) idealized the gouge to be a medium formed by circular disk-shaped pieces filling the space between two planes acting like bearings. Here, we present a more realistic approximation, considering that the surfaces of the tectonic plates are irregular and the space between them contains fragments of very diverse shape.

The abundance of models that manifest the catastrophic mechanical nature of earthquakes is expected since the mechanism of earthquake triggering, although not well known, reveals itself as a catastrophic one. The irregularity of the profiles of the tectonic plates and fault planes has been pointed out as a main cause of earthquakes and in [12] the Gutenberg-Richter law for large earthquakes was obtained on the basis of assuming a Brownian shape of the profiles and the hypothesis that energy release is proportional to the overlap interval between profiles. In other models [13] the material between the fault planes is considered. Nevertheless, the irregularities of the fault planes can be combined with the distribution of fragments between them to develop a mechanism of triggering earthquakes, which can be essentially the same as in [13] owing to the irregular shape of the profiles and fragments.

Let us start from the situation illustrated in the figure 2: two irregular profiles (no predetermined shape is assumed) are able to slip as shown in the figure. Stress in the structure accumulates until one of the asperities is broken. Then, the slip occurs. In the other hand, we can considerer the phenomena in a small scale, that is, we are showing now the details of the gouge that can produce small earthquakes (figure 3). The motion can be hindered not only by the overlapping of two irregularities of the profiles, but also by the eventual relative position of several fragments as illustrated in the figure between the points "a" and "b". Stress in the resulting structure accumulates until a displacement of one of the asperities or even its breakage in the point of contact with the gouge fragment leads to a relative displacement of the fault planes of the order of the size of the hindering fragment "r". Then, the eventual release of stress, whatever be the cause, leads to such a displacement with the subsequent liberation of energy. We assume this energy "$\varepsilon$" to be proportional to "r", so that the size distribution of the fragments in the gouge can reflect the energy distribution of earthquakes generated by this mechanism.

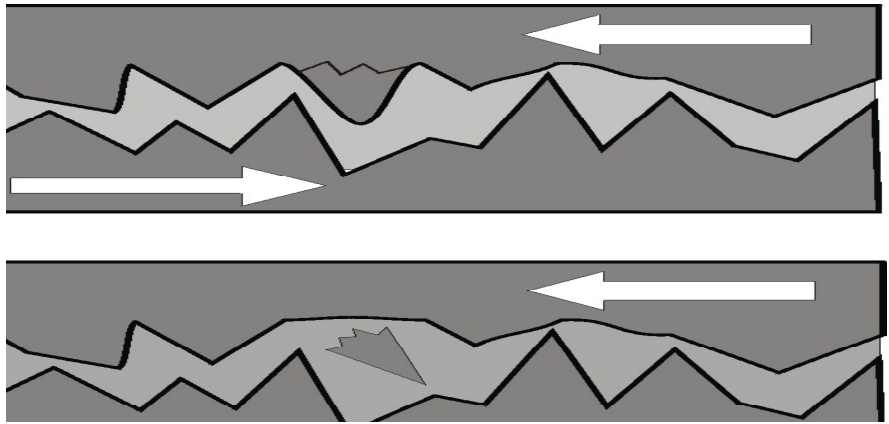

Fig. 2. – Interaction of irregular profiles

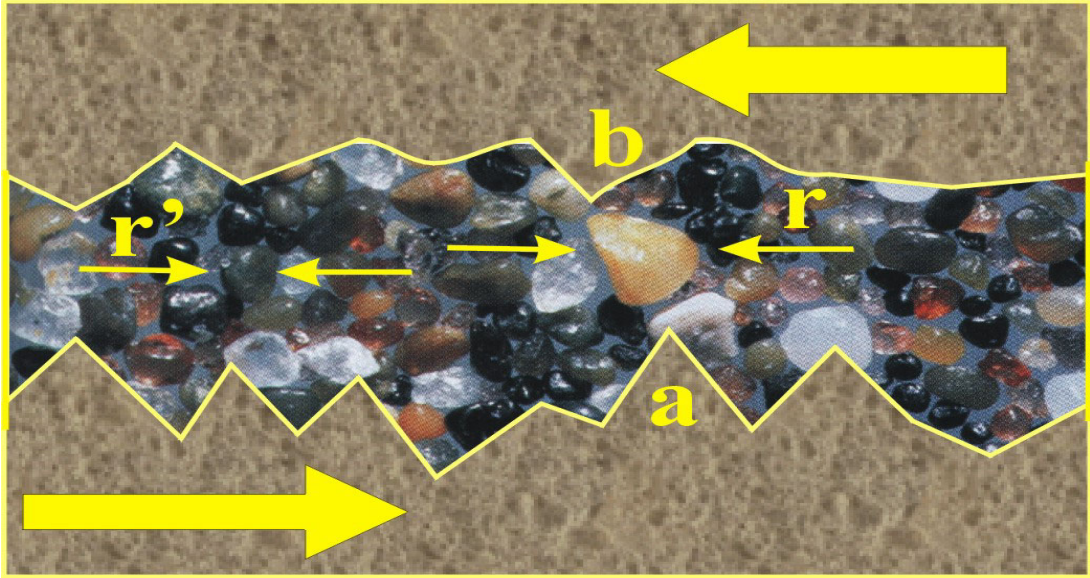

Fig3. – Profiles interacting through fragments

As already pointed out, the size distribution function of fragments between the fault planes should reflect the energy distribution function of the earthquakes. We can assume that the constant interaction and local breakage of the fault planes produce the fragments. The process of fault slip can be considered to occur in a homogeneous fashion in all the deep of the fault so that in any plane transverse to the depth of the fault the situation is the same. To deduce the size distribution function of the fragments we consider a two-dimensional frame as the one illustrated in figure 2 or 3. Then, our problem is to find the distribution of fragments by area.

As fractioning is a paradigm of non-extensivity, it may be necessary to use non-extensive statistics instead of the Boltzmann-Gibbs one to describe the size distribution function of the fragments. As in the previous section, we will apply the maximum entropy principle for the Tsallis entropy.

The Tsallis entropy for our problem has the form:

$$S_q = k \frac{1 - \int p^q(\sigma) d\sigma}{q - 1} \qquad (16)$$

where $p(\sigma)$ is the probability of finding a fragment of relative surface $\sigma$ referred to a characteristic surface of the system, and $q$ is a real number. $k$ is Boltzmann's constant. It is easy to see that this entropy reduces to the Boltzmann's one when $q \to 1$. The sum by all states in the entropy is here expressed through the integration in all sizes of the fragments up to the largest surface $\Sigma$ of the collection of fragments. As we already know, the Tsallis formulation involves the introduction of at least two constraints. The first one is the normalization of $p(\sigma)$:

$$\int_0^\Sigma p(\sigma)d\sigma = 1 \qquad (17)$$

and the other is the "*ad hoc*" condition about the q-mean value, which in our case can be expressed as:

$$\int_0^\Sigma \sigma\, p^q(\sigma)d\sigma = <<\sigma>>_q \qquad (18)$$

This condition reduces to the definition of the mean value when $q \to 1$. More information concerning the constraints that can be imposed in the formulation can be seen in [8]. As we already said, fracture is a paradigm of such long-range interaction phenomenon, then the problem is to find the extremum of $\dfrac{S_q}{k}$ subject to the conditions given by formulas 15 and 16. To simplify we will assume $<<\sigma>>_q = 1$ (we will see that this has no effect in the final result). We define the lagrangian function $\Gamma$ to be:

$$\Gamma = \frac{S_q}{k} + \lambda \int_0^\Sigma p(\sigma)d\sigma + \beta \int_0^\Sigma \sigma p^q(\sigma)d\sigma \qquad (19)$$

with $\lambda$ and $\beta$ the Lagrange multipliers. Application of the method of Lagrange multipliers follows with:

$$\frac{\partial \Gamma}{\partial p} = 0 \qquad (20)$$

and we find that:

$$p(\sigma)\,d\sigma = \frac{C_1 d\sigma}{[(1-C_2)\sigma]^{1/(q-1)}} \qquad (21)$$

where $C_1$ and $C_2$ are constants to be determined under the conditions of equation 2 and 3; if we do that, it is possible to find:

$$p(\sigma)\,d\sigma = \frac{(2-q)^{1/(2-q)}\,d\sigma}{\left[1+(q-1)(2-q)^{1/(2-q)}\sigma\right]^{1/(q-1)}} \qquad (22)$$

If we now introduce the proportionality of the released relative energy $\varepsilon$ with the linear dimension r of the fragments, as $\sigma$ scales with $r^2$, the resulting expression for the energy distribution function of the earthquakes due to this mechanism is:

$$p(\varepsilon)\,d\varepsilon = \frac{2C_1 k\varepsilon\, d\varepsilon}{\left[1 + C_2 k\,\varepsilon^2\right]^{1/(q-1)}} \qquad (23)$$

where the probability of the energy $p(\varepsilon) = n(\varepsilon)/N$ being $n(\varepsilon)$ the number of earthquakes of energy ε and N the total number of earthquakes. Hence, we have obtained an analytic expression which can be fitted to the energy distribution of earthquakes through $k$ and $q$. This was obtained from a simple model starting from first principles. No *ad hoc* hypothesis was introduced but the proportionality of "ε" and "r", which seems justified. A similar treatment can be performed with Boltzmann's entropy.

To use the common frequency-magnitude distribution, the cumulative number was calculated as the integral from "ε" to "∞"of the formula 8; then:

$$N(\varepsilon >) \;=\; \int_{\varepsilon}^{\infty} p(\varepsilon)\,d\varepsilon \qquad (24)$$

On the other hand $m \propto \log(\varepsilon)$ where m is the magnitude, so we get:

$$\log(N(>m)) = \log N + \left(\frac{2-q}{1-q}\right)\log\left[1 + k(q-1)(2-q)^{\frac{1-q}{q-2}} \cdot 10^{2m}\right] \qquad (25)$$

This is not a trivial result because it means that $1 < q < 2$ and we can interpret these limitations for q attending to the range of spatial correlations between the interacting elements in the problem. On the other hand, use of Boltzmann's entropy following the same method leads to:

$$\log(N(>m)) = a - b \cdot 10^m \qquad (26)$$

Both laws are different from the Gutenberg-Richter law:

$$\log(N(>m)) = a - bm \qquad (27)$$

### Test for small earthquakes

The Andalusian Institute of Geophysics and Seismic Disaster Prevention compiled the earthquake catalogue used in this study. The earthquakes were observed by the Andalusian Seismic Network of the Andalusian Institute of Geophysics and Seismic Disaster Prevention that consists of more than 20 observational stations [15]. The analyzed area is the rectangular region between 35º and 38º North Latitude and between 0º and 5º West Longitude, hereinafter called the South of the Iberian Peninsula region. The epicenters of the earthquakes of the South of the Iberian Peninsula region are registered in a catalogue of more than 20000 events. Error of the hypocenters' location in the x, y and z directions are more or less ±1 km, ±1 km and ±2 km, respectively . The seismicity during the period from 1985 to 2000 may be considered normal, i.e., without major seismic events. The Gutenberg-Richter relation is satisfied in this data set, for earthquakes with magnitude greater than 2.5. The data is assumed to be free of observational bias and abnormal seismicity. Boltzmann's fit and Gutenberg-Richter's fit are in figure 4 and Tsallis's fit is in figure5. The assumptions applied to the Boltzmann formulation does not work for earthquakes of large magnitude. The frequency of event occurrence is a log-linear function for the range of magnitude between 2.5 and 5; this is the Gutenberg-Richter law and, as we expected, works correctly for moderate seismicity in the region. Finally, Tsallis's fit reflect that our assumptions for the gouge earthquakes are correct because the curve fits very well for seismicity ranging from 0 to 2.5 magnitude and also for higher magnitude Tsallis expression gives a good fitting for all the magnitude values. The obtained q value has a value equal to 1,645 with a correlation factor equal to 0,998.

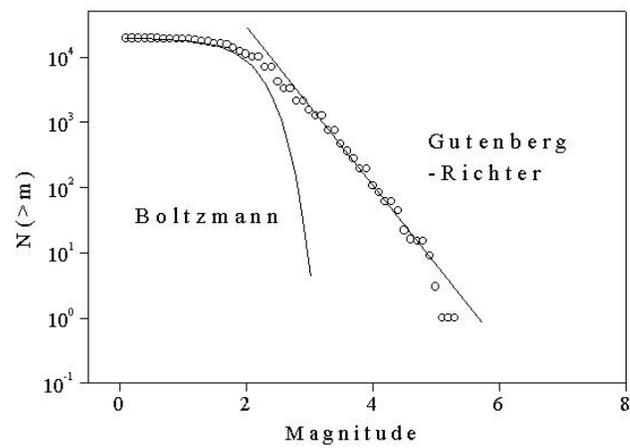

Fig4. - fitting of the Andalusian data with formula 26 and with the Gutenberg-Richter law, formula 27

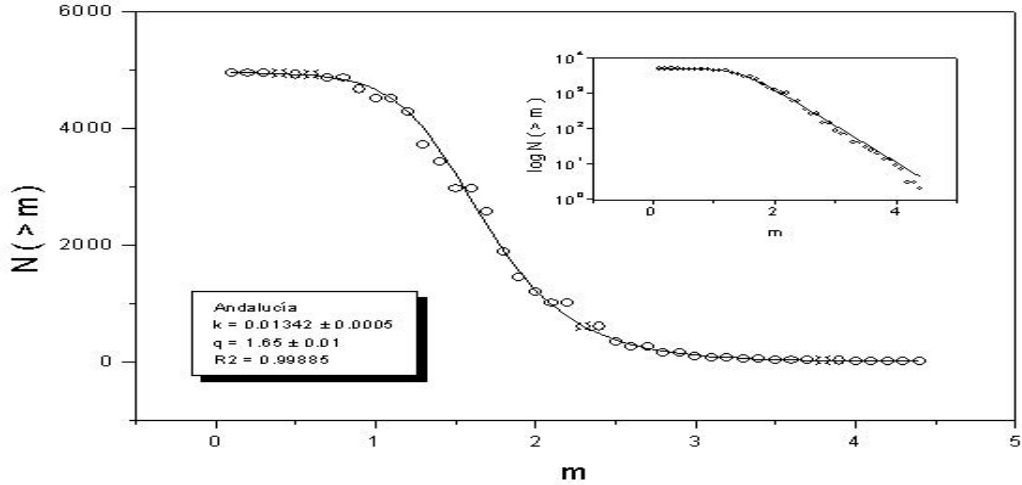

Fig5.- Fitting of the Andalusian data with formula 25. Logarithmic plot is shown in the insert

## Test for large earthquakes

Lomnitz and Lomnitz have proposed a stochastic model of strain accumulation and release at plate boundaries [16] . The model leads to a generalized Gutenberg-Richter relation in terms of *G(m)* the cumulative excedence of a magnitude *m*. The relation tends to the original Gutenberg-Richter relation in the low magnitude range and at high magnitudes it provides estimates of the probability occurrence which are significantly more realistic than the Gutenberg-Richter law. They have obtained excellent agreement with the data of the Chinese earthquake catalogue covering a time period of 2753 years. This catalogue contains the earthquakes for a threshold magnitude *m = 6.0*; it is the longest published catalog of historical earthquakes in any region. Lomnitz and Lomnitz (1979) excedence define like:

$$G(\varepsilon) = \int_{\varepsilon}^{\infty} p(\varepsilon)d\varepsilon \qquad (28)$$

and by using our formula (25) we have fit the Chinese catalogue and results can be seen in FIGURE 6. As we can observe, in this case q = 1,688 and the correlation factor is 0.994.

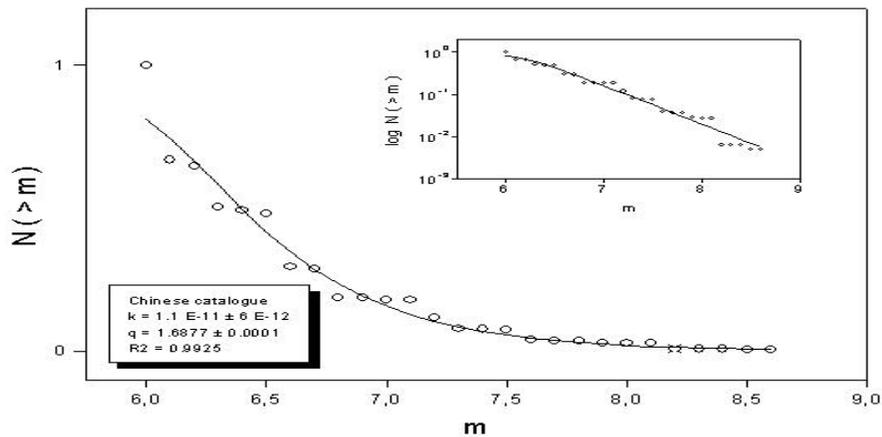

Fig 6. - Fitting of the chinese catalogue using formula 25. Logarithmic plot is shown in the insert. As previously, the results are pretty good

## Results for a whole catalogue

Two different catalogues are used in this section. First, a large catalogue from the United States Geological Survey. The catalogue includes all the earthquakes in the California area, that is, all the San Andreas fault systems; the temporal period is from 1990 to actuality. More than 500000 earthquakes were processed and the results are in FIGURE 7. As we can see, Tsallis formulation can fit all the earthquakes in the catalogue. The value of q is 1,676 and the correlation factor is 0.999. The second large catalogue is from the Geographic National Institute (Spain) and it has all the seismic data in the Iberian Peninsula (Spain). More than 10000 earthquakes are collected between 1970 to actuality. The results are shown in the FIGURE 8. Tsallis formulation can fit also this data; the results are q = 1,663 and the correlation factor 0,999.

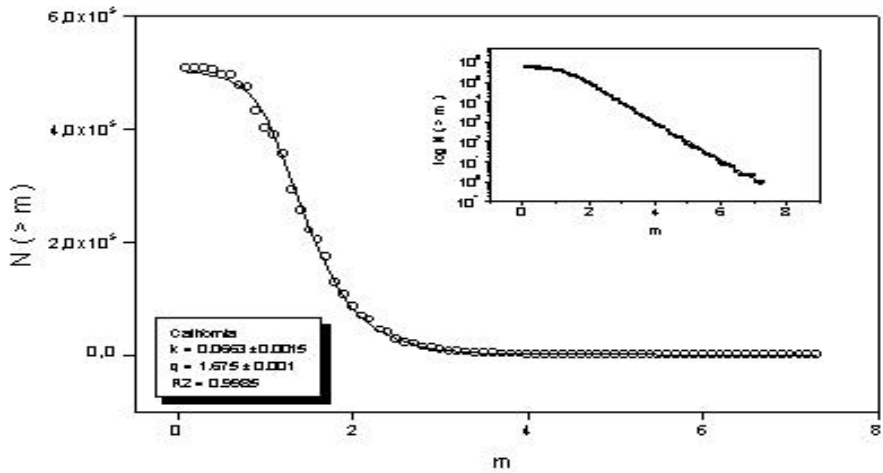

Fig 7. -Results for the catalogue of the San Andreas fault. The legend is the same as for previous figures

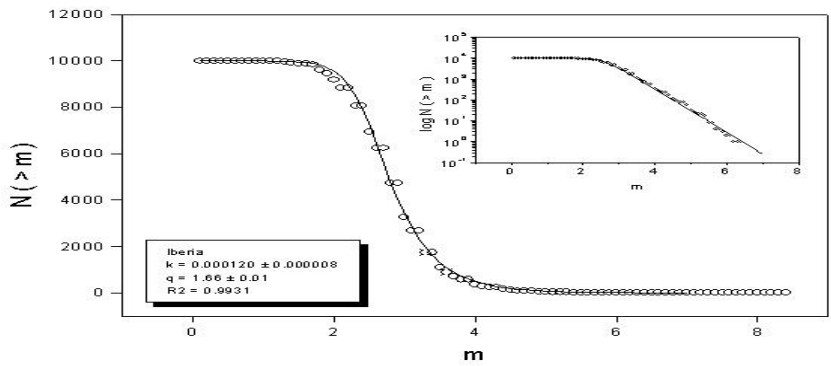

Fig 8. -Results for the catalogue of the Iberian Peninsula. The legend is the same as for previous figures.

## *Conclusions*

Fragments and fragmentation processes are present in an enormous amount of natural processes, and frequently this aspect does not receive adequate attention. With the reviewed examples we try to highlight two main aspects to take into account in the science of complex phenomena: One is that frequently a simple geometrical viewpoint is able to clarify problems that are intractable by other methods. So, fractal geometry and the theory of Levy distributions are called to play an important role in the physics of complex phenomena. The second aspect is the importance of a generalization of many concepts originated in the statistical physics of equilibrium systems, better to say the "orthodox" statistical physics; This generalization is a promising way to get a formulation of many problems that, up to now, have been poorly studied due to the inadequacy of current methods. In this paper, with the results shown in fragmentation and earthquakes, Tsallis entropy showed to be a promising tool to start in this way

*References*

1. - M. Matsushita *J. Phys. Soc. Jpn* **54**, 857 (1984).
2. - R. Englman, N. Rivier and Z. Jaeger *Phil. Mag* **56**, 751 (1987).
3. -O. Sotolongo, E. López, F. Barrera, J Marín *Phys. Rev. E* **49**. 4027 (1994)PRE
4. - O. Sotolongo-Costa, and R. Grau Crespo *Rev. Mex. Fís* **44** (5) 461 (1998).
5. - T. Ishii, M. Matsushita *J. Phys. Soc. Jpn* **61**, 3474 (1992).
6. - O. Soto0longo-Costa, Y. Moreno J.J. Lloveras, J.C. Antoranz *Phys. Rev Lett*. **76**, 42 (1996).
7. - X. Li, R.S. Tankin. *Combust. Sci and Tech*. **50**, 65 (1987).
8. – C.Tsallis. *Brazilian J. Phys* **29**, 1 (1999).
9. – C.Tsallis *J. Stat. Phys* **52**,479 (1988).
10. - Burridge, R. and L. Knopoff. *Bull. Seism. Soc. Am*. **57**, 341. (1967)
11.- Olami, H; J. S. Feder and K. Christensen;; Self-organized criticality in a continuous, nonconservative cellular automaton modeling earthquakes. *Phys. Rev. Lett*. **68**, 1244. (1992).
12. -De Rubeis, V.; R. Hallgas, V. Loreto, G. Paladin, L. Pietronero and P. Tosi. *Phys. Rev. Lett*. **76**, 2599. (1996).
13. -Herrmann, H.J.; G. Mantica and D. Bessis. *Phys. Rev. Lett*. **65**, 3223. (1990).
14, - Sornette, D. *Phys. Reports* vol **313:5**, 328-292. .(1999).
15. - Posadas, A.M.; F. Vidal; F. De Miguel; G. Alguacil; J. Peña; J.M. Ibañez; J. Morales; *J. Geophys. Res*. **98**, B2, 1923-1932.( 1989).
16. - Lomnitz-Adler, J; Lomnitz, C. *Bull. Seism. Soc. Am.,* **69**, 4 1209-1214 (1979).